%% file: 0_CQA.tex
\documentclass[sigconf]{acmart}
 \usepackage{textcomp, libertine}
\usepackage{booktabs} 
\usepackage[normalem]{ulem}
\usepackage{multirow}
\usepackage{bm}
\useunder{\uline}{\ul}{}
\usepackage{subcaption}
\usepackage[para]{footmisc}
\usepackage{enumitem}
\usepackage[ruled]{algorithm2e}
\usepackage{amssymb}
\usepackage{dsfont}
\hypersetup{
,bookmarksnumbered=true%
,hypertexnames=false%
,breaklinks=true%
,colorlinks=true%
,linkcolor=black%
,citecolor=black%
,urlcolor=black%
}

\definecolor{tealblue}{rgb}{0.21, 0.46, 0.53}
\definecolor{wildstrawberry}{rgb}{1.0, 0.26, 0.64}
\definecolor{ao(english)}{rgb}{0.0, 0.5, 0.0}

\def\ignore#1{}

\setcopyright{acmcopyright} 
\begin{document}
\fancyhead{} 
\title{Attentive History Selection for Conversational \\Question Answering}

\author{Chen Qu$^1$ \quad Liu Yang$^1${\footnotemark[1]} \quad Minghui Qiu$^2$ \quad Yongfeng Zhang$^3$ \quad Cen Chen$^4$ \quad\\ W. Bruce Croft$^1$ \quad Mohit Iyyer$^1$}

\affiliation{%
	\institution{
		$^1$ University of Massachusetts Amherst \quad
		$^2$ Alibaba Group \quad
        $^3$ Rutgers University \quad
        $^4$ Ant Financial Services Group} 
}
\email{{chenqu, lyang, croft, miyyer}@cs.umass.edu, minghui.qmh@alibaba-inc.com} \email{yongfeng.zhang@rutgers.edu,chencen.cc@antfin.com}

\begin{abstract}
Conversational question answering (ConvQA) is a simplified but concrete setting of conversational search~\cite{hae}. One of its major challenges is to leverage the conversation history to understand and answer the current question. 
In this work, we propose a novel solution for ConvQA that involves three aspects. First, we propose a \textit{positional history answer embedding} method to encode conversation history with position information using BERT~\cite{bert} in a natural way. BERT is a powerful technique for text representation. 
Second, we design a \textit{history attention mechanism} (HAM) to conduct a ``soft selection'' for conversation histories. This method attends to history turns with different weights based on how helpful they are on answering the current question. Third, in addition to handling conversation history, we take advantage of \textit{multi-task learning} (MTL) to do answer prediction along with another essential conversation task (dialog act prediction) using a uniform model architecture.
MTL is able to learn more expressive and generic representations to improve the performance of ConvQA. We demonstrate the effectiveness of our model with extensive experimental evaluations on QuAC, a large-scale ConvQA dataset. We show that position information plays an important role in conversation history modeling. We also visualize the history attention and provide new insights into conversation history understanding. 
\end{abstract}


\settopmatter{printacmref=false, printfolios=false}


\copyrightyear{2019} 
\acmYear{2019} 
\acmConference[CIKM '19]{The 28th ACM International Conference on Information and Knowledge Management}{November 3--7, 2019}{Beijing, China}
\acmBooktitle{The 28th ACM International Conference on Information and Knowledge Management (CIKM '19), November 3--7, 2019, Beijing, China}
\acmPrice{15.00}
\acmDOI{10.1145/3357384.3357905}
\acmISBN{978-1-4503-6976-3/19/11}

\maketitle

\renewcommand*{\thefootnote}{\fnsymbol{footnote}}
\footnotetext[1]{Liu Yang is at Google currently.}
\renewcommand*{\thefootnote}{\arabic{footnote}}

\input{1_introduction}

\input{2_related_work}
\input{3_our_approach}

\input{4_experiments}
\input{5_conclusions}

\begin{acks}
This work was supported in part by the Center for Intelligent Information Retrieval and in part by NSF IIS-1715095. Any opinions, findings and conclusions or recommendations expressed in this material are those of the authors and do not necessarily reflect those of the sponsor.
\end{acks}

\bibliographystyle{ACM-Reference-Format}
\bibliography{acmart} 

\end{document}

%% file: 1_introduction.tex
{\fontsize{8pt}{8pt} \selectfont
\textbf{ACM Reference Format:}\\
Chen Qu, Liu Yang, Minghui Qiu, Yongfeng Zhang, Cen Chen, W. Bruce Croft, and Mohit Iyyer. 2019. Attentive History Selection for Conversational Question Answering. In \textit{The 28th ACM International Conference on Information and Knowledge Management (CIKM '19), November 3--7, 2019, Beijing, China.} ACM, New York, NY, USA, 10 pages. \url{https://doi.org/10.1145/3357384.3357905}}

\section{Introduction}
\label{sec:intro}

It has been a longstanding goal in the information retrieval (IR) community to design a search system that can retrieve information in an interactive and iterative manner~\cite{Belkin1994CasesS,i3r,Oddy1977Information,Kotov2010TowardsNQ}. With the rapid development of artificial intelligence and conversational AI~\cite{Gao2018NeuralAT}, IR researchers have begun to explore a concrete implementation of this research goal, referred to as conversational search. Contributions from both industry and academia have greatly boosted the research progress in conversational AI, resulting in a wide range of personal assistant products. Typical examples include Apple Siri, Google Assistant, Amazon Alexa, and Alibaba AliMe~\cite{Li2017AliMeA}. An increasing number of users are relying on these systems to finish everyday tasks, such as setting a timer or placing an order. Some users also interact with them for entertainment or even as an emotional companion. Although current personal assistant systems are capable of completing tasks and even conducting smalltalk, they cannot handle information-seeking conversations with complicated information needs that require multiple turns of interaction. Conversational personal assistant systems serve as an appropriate media for interactive information retrieval, but much work needs to be done to enable functional conversational search via such systems. 

A typical conversational search process involves multiple ``cycles''~\cite{hae}. In each cycle, a user first specifies an information need and then an agent (a system) retrieves answers iteratively either based on the user's feedback or by asking for missing information proactively~\cite{Zhang2018TowardsCS}. The user could ask a follow-up question and shift to a new but related information need, entering the next cycle of conversational search. Previous work~\cite{hae} argues that conversational question answering (ConvQA) is a simplified but concrete setting of conversational search. Although the current ConvQA setting does not involve asking spontaneously, it is a tangible task for researchers to work on modeling the change of information needs across cycles. Meanwhile, conversation history plays an important role in understanding the latest information need and thus is beneficial for answering the current question. For example, we show that coreferences are common across conversation history in Table~\ref{tab:example}. Therefore, one of the major focuses of this work is handling conversation history in a ConvQA setting. 

\begin{table}[htbp]
\caption{An example of an information-seeking dialog from QuAC. ``R'', ``U'', and ``A'' denote role, user, and agent. Co-references and related terms are marked in the same color across history turns. $\text{Q}_2$, $\text{Q}_4$, $\text{Q}_5$ and $\text{Q}_6$ are closely related to their immediate previous turn(s) while $\text{Q}_7$ is related to a remote question $\text{Q}_1$. Also, $\text{Q}_3$ does not follow up on $\text{Q}_2$ but shifts to a new topic. This table is best viewed in color.
}
\label{tab:example}
\footnotesize
\tabcolsep=0.08cm
\vspace{-0.4cm}
\begin{tabular}{@{}clcl@{}}
\toprule
\multicolumn{4}{l}{Topic: Lorrie Morgan's music career}                                                                                                             \\ \midrule
\multicolumn{1}{c|}{\#}                 & \multicolumn{1}{l|}{ID}  & \multicolumn{1}{c|}{R} & \multicolumn{1}{c}{Utterance}                                            \\ \midrule
\multicolumn{1}{c|}{\multirow{2}{*}{1}} & \multicolumn{1}{l|}{$\text{Q}_1$}  & \multicolumn{1}{c|}{U}     & What is relevant about Lorrie's \textcolor{purple}{musical career? }                                      \\
\multicolumn{1}{c|}{}                   & \multicolumn{1}{l|}{$\text{A}_1$}  & \multicolumn{1}{c|}{A}     & ... her first \textcolor{brown}{album} on that label, \textcolor{brown}{Leave the Light On}, was released in 1989.         \\ \midrule
\multicolumn{1}{c|}{\multirow{2}{*}{2}} & \multicolumn{1}{l|}{$\text{Q}_2$}  & \multicolumn{1}{c|}{U}     & What songs are included in the \textcolor{brown}{album}?                                            \\
\multicolumn{1}{c|}{}                   & \multicolumn{1}{l|}{$\text{A}_2$}  & \multicolumn{1}{c|}{A}     & CANNOTANSWER    \\ \midrule
\multicolumn{1}{c|}{\multirow{2}{*}{3}} & \multicolumn{1}{l|}{$\text{Q}_3$}  & \multicolumn{1}{c|}{U}     & Are there any other interesting aspects about this article?                                       \\
\multicolumn{1}{c|}{}                   & \multicolumn{1}{l|}{$\text{A}_3$}  & \multicolumn{1}{c|}{A}     & made \textcolor{orange}{her first appearance} on the Grand Ole Opry at age 13,         \\ \midrule
\multicolumn{1}{c|}{\multirow{2}{*}{4}} & \multicolumn{1}{l|}{$\text{Q}_4$}  & \multicolumn{1}{c|}{U}     & What did she do after \textcolor{orange}{her first appearance}?                                            \\
\multicolumn{1}{c|}{}                   & \multicolumn{1}{l|}{$\text{A}_4$}  & \multicolumn{1}{c|}{A}     & ... she took over his \textcolor{blue}{band} at age 16 and began leading the \textcolor{blue}{group} ...    \\ \midrule
\multicolumn{1}{c|}{\multirow{2}{*}{5}} & \multicolumn{1}{l|}{$\text{Q}_5$}  & \multicolumn{1}{c|}{U}     & What important work did she do with the \textcolor{blue}{band}?                                       \\
\multicolumn{1}{c|}{}                   & \multicolumn{1}{l|}{$\text{A}_5$}  & \multicolumn{1}{c|}{A}     & leading the \textcolor{blue}{group} through various club gigs.         \\ \midrule
\multicolumn{1}{c|}{\multirow{2}{*}{6}} & \multicolumn{1}{l|}{$\text{Q}_6$}  & \multicolumn{1}{c|}{U}     & What songs did she played with the \textcolor{blue}{group}?                                            \\
\multicolumn{1}{c|}{}                   & \multicolumn{1}{l|}{$\text{A}_6$}  & \multicolumn{1}{c|}{A}     & CANNOTANSWER    \\ \midrule
\multicolumn{1}{c|}{\multirow{2}{*}{7}} & \multicolumn{1}{l|}{$\text{Q}_7$}  & \multicolumn{1}{c|}{U}     & What are other interesting aspects of her \textcolor{purple}{musical career}?                                            \\
\multicolumn{1}{c|}{}                   & \multicolumn{1}{l|}{$\text{A}_6$}  & \multicolumn{1}{c|}{A}     & \textit{To be predicted ... }   \\ \bottomrule

\end{tabular}
\end{table}

In two recent ConvQA datasets, QuAC~\cite{quac} and CoQA~\cite{coqa}, ConvQA is formalized as an answer span prediction problem similar in SQuAD~\cite{squad,squad2}. Specifically, given a question, a passage, and the conversation history preceding the question, the task is to predict a span in the passage that answers the question. In contrast to typical machine comprehension (MC) models, it is essential to handle conversation history in this task. Previous work~\cite{hae} introduced a general framework to deal with conversation history in ConvQA, where a history selection module first selects helpful history turns and a history modeling module then incorporates the selected turns. In this work, we extend the same concepts of history selection and modeling with a fundamentally different model architecture.

On the aspect of history selection, existing models~\cite{quac,coqa} select conversation history with a simple heuristic that assumes immediate previous turns are more helpful than others. This assumption, however, is not necessarily true. \citet{Yatskar2018AQC} conducted a qualitative analysis on QuAC by observing 50 randomly sampled passages and their corresponding 302 questions. He showed that 35.4\% and 5.6\% of questions have the dialog behaviors of \textit{topic shift} and \textit{topic return} respectively. A topic shift suggests that the current question shifts to a new topic, such as the $\text{Q}_3$ in Table~\ref{tab:example}. While topic return means that the current question is about a topic that has previously been shifted away from. For example, $\text{Q}_7$ returns to the same topic in $\text{Q}_1$ in Table~\ref{tab:example}. In both cases, the current question is not directly relevant to immediate previous turns. It could be unhelpful or even harmful to always incorporate immediate previous turns. Although we expect this heuristic to work well in many cases where the current question is \textit{drilling down} on the topic being discussed, it might not work for topic shift or topic return. There is no published work that focuses on \textit{learning} to select or re-weight conversation history turns. To address this issue, we propose a \textit{history attention mechanism} (HAM) that learns to attend to all available history turns with different weights. This method increases the scope of candidate histories to include remote yet potentially helpful history turns. Meanwhile, it promotes useful history turns with large attention weights and demotes unhelpful ones with small weights. More importantly, the history attention weights provide explainable interpretations to understand the model results and thus can provide new insights in this task.

In addition, on the aspect of history modeling, some existing methods either simply prepend the selected history turns to the current question~\cite{coqa,sdnet} or use complicated recurrent structures to model the conversation history~\cite{flowqa}, generating relatively large system overhead. Another work~\cite{hae} introduces a history answer embedding (HAE) method to incorporate the conversation history to BERT in a natural way. However, they fail to consider the position of a history utterance in the dialog. Since the utility of a history utterance could be related to its position, we propose to consider the position information in HAE, resulting in a \textit{positional history answer embedding} (PosHAE) method. We show that position information plays an important role in conversation history modeling.

Furthermore, we introduce a new angle to tackle the problem of ConvQA.
We take advantage of \textit{multi-task learning} (MTL) to do answer span prediction along with another essential conversation task (dialog act prediction) using a uniform model architecture.
Dialog act prediction is necessary in ConvQA systems because dialog acts can reveal crucial information about user intents and thus help the system provide better answers. More importantly, by applying this multi-task learning scheme, the model learns to produce more generic and expressive representations~\cite{Liu2019MultiTaskDN}, due to additional supervising signals and the regularization effect when optimizing for multiple tasks. We show that these benefits have contributions to the model performance for the dialog action prediction task.

In this work, we propose a novel solution to tackle ConvQA. 
We boost the performance from three different angles, i.e., history selection, history modeling, and multi-task learning. Our contributions can be summarized as follows:

\begin{enumerate}[leftmargin=1em]
    \item To better conduct history selection,
    we introduce a history attention mechanism to conduct a ``soft selection'' for conversation histories. This method attends to history turns with different weights based on how helpful they are on answering the current question. This method enjoys good explainability and can provide new insights to the ConvQA task.
    
    \item To enhance history modeling, we incorporate the history position information into history answer embedding~\cite{hae}, resulting in a positional history answer embedding method. Inspired by the latest breakthrough in language modeling, we leverage BERT to jointly model the given question, passage and conversation history, where BERT is adapted to a conversation setting.
    
    \item To further improve the performance of ConvQA, we jointly learn answer span prediction and dialog act prediction in a multi-task learning setting. We take  advantage of MTL to learn more generalizable representations. 
    
    \item We conduct extensive experimental evaluations to demonstrate the effectiveness of our model and to provide new insights for the  ConvQA task. The implementation of our model has been open-sourced to the research community.\footnote{\url{https://github.com/prdwb/attentive_history_selection}}
    
\end{enumerate}

%% file: 2_related_work.tex
\section{Related Work}
\label{sec:relatedwork}
Our work is closely related to several research areas, including machine comprehension, conversational question answering, conversational search, and multi-task learning.

\textbf{Machine Comprehension}. Machine reading comprehension is one of the most popular tasks in natural language processing. Many high-quality challenges and datasets~\cite{squad,squad2,Marco,TriviaQA,GoogleNQ} have greatly boosted the research progress in this field, resulting in a wide range of model architectures~\cite{bidaf,Hu2018ReinforcedMR,Wang2017GatedSN,Huang2017FusionNetFV,Clark2018SimpleAE}. One of the most influential datasets in this field is SQuAD (The Stanford Question Answering Dataset)~\cite{squad,squad2}. The reading comprehension task in SQuAD is conducted in a single-turn QA manner. The system is given a passage and a question. The goal is to answer the question by predicting an answer span in the passage. Extractive answers in this task enable easy and fair evaluations compared with other datasets that have abstractive answers generated by human. The recently proposed BERT~\cite{bert} model pre-trains language representations with bidirectional encoder representations from transformers and achieves exceptional results on this task. BERT has been one of the most popular base models and testbeds for IR and NLP tasks including machine comprehension.

\textbf{Conversational Question Answering}.
CoQA~\cite{coqa} and QuAC~\cite{quac} are two large-scale ConvQA datasets. The ConvQA task in these datasets is very similar to the MC task in SQuAD. A major difference is that the questions in ConvQA are organized in conversations. Although both datasets feature ConvQA in context, they come with very different properties. Questions in CoQA are often factoid with simple entity-based answers while QuAC consists of mostly non-factoid QAs. More importantly, information-seekers in QuAC have access to the title of the passage only, simulating an information need. QuAC also comes with dialog acts, which is an essential component in this interactive information retrieval process. The dialog acts provide an opportunity to study the multi-task learning of answer span prediction and dialog act prediction. Overall, the information-seeking setting in QuAC is more in line with our interest since we are working towards the goal of conversational search. Thus, we focus on QuAC in this work. Although leaderboards of CoQA\footnote{\url{https://stanfordnlp.github.io/coqa/}} and QuAC\footnote{\url{http://quac.ai/}} show more than two dozen submissions, these models are mostly work done in parallel with ours and rarely have descriptions, papers, or codes.  

Previous work~\cite{hae} proposed a ``history selection - history modeling'' framework to handle conversation history in ConvQA. In terms of history selection, existing works\cite{quac,coqa,sdnet,flowqa,hae} adopt a simple heuristic of selecting immediate previous turns. This heuristic, however, does not work for complicated dialog behaviors. There is no published work that focuses on \textit{learning} to select or re-weight conversation history turns. To address this issue, we propose a history attention mechanism, which is a learned strategy to attend to history turns with different weights according to how helpful they are on answering the current question. In terms of history modeling, existing methods simply prepend history turns to the current question~\cite{coqa,sdnet} or use a recurrent structure to model the representations of history turns~\cite{flowqa}, which has a lower training efficiency~\cite{hae}. Recently, a history answer embedding method~\cite{hae} was proposed to learn two unique embeddings to denote whether a passage token is in history answers. However, this method fails to consider the position information of history turns. We propose to enhance this method by incorporating the position information into the history answer embeddings. 

\textbf{Conversational Search}.
Conversational search is an emerging topic in the IR community, however, the concept of it dates back to several early works~\cite{Belkin1994CasesS,i3r,Oddy1977Information}. Conversational search poses unique challenges as answers are retrieved in an iterative and interactive manner. Much effort is being made towards the goal of conversational search. The emerging of neural networks has made it possible to train conversation models in an end-to-end manner. Neural approaches are widely used in various conversation tasks, such as conversational recommendation~\cite{Zhang2018TowardsCS}, user intent prediction~\cite{UserIntentPred}, next question prediction~\cite{Yang2017NeuralMM}, and response ranking~\cite{Yang2018ResponseRW,Guo2019ADL}. In addition, researchers also conduct observational studies~\cite{Qu2018AnalyzingAC,Chuklin2018ProsodyMF,Trippas2018InformingTD,misc,answer_interaction} to inform the design of conversational search systems. In this work, we focus on handling conversation history and using a multi-task learning setting to jointly learn dialog act prediction and answer span prediction. These are essential steps towards the goal of building functional conversational search systems.

\textbf{Multi-task Learning}. Multi-tasking learning has been a widely used technique to learn more powerful representations with deep neural networks~\cite{Zhang2018ASO}. A common paradigm is to employ separate task-specific layers on top of a shared encoder~\cite{Liu2019MultiTaskDN,Liu2015RepresentationLU,Xu2018MultiTaskLF}. The encoder is able to learn representations that are more expressive, generic and transferable. Our model also adopts this paradigm. Not only can we enjoy the advantages of MTL, but also handle two essential tasks in ConvQA, answer span prediction and dialog act prediction, with a uniform model architecture.

%% file: 3_our_approach.tex
\section{Our Approach}
\label{sec:our-approach}
\subsection{Task Definition}
\label{subsec:task}
The ConvQA task is defined as follows~\cite{quac,coqa}. Given a passage $p$, the $k$-th question $q_{k}$ in a conversation, and the conversation history $\mathbf{H}_k$ preceding $q_k$, the task is to answer $q_{k}$ by predicting an answer span $a_k$ within the passage $p$. The conversation history $\mathbf{H}_k$ contains $k-1$ turns, where the $i$-th turn $\mathbf{H}_k^i$ contains a question $q_i$ and its groundtruth answer $a_i$. Formally, $\mathbf{H}_k = \{(q_i, a_i)\}_{i=1}^{k-1}$. One of the unique challenges of ConvQA is to leverage the conversation history to understand and answer the current question.

Additionally, an important task relevant to conversation modeling is dialog act prediction. QuAC~\cite{quac} provides two dialog acts, namely, \textit{affirmation} (Yes/No) and \textit{continuation} (Follow up). The affirmation dialog act $v^a$ consists of three possible labels: \{\texttt{yes, no, neither}\}. The continuation dialog act $v^c$ also consists of three possible labels: \{\texttt{follow up, maybe follow up, don't follow up}\}. Each question is labeled with both dialog acts. The labels for each dialog act are mutually exclusive. This dialog act prediction task is essentially two sentence classification tasks. Therefore, a complete training instance is composed of the model input $(q_{k}, p, \mathbf{H}_k)$ and its ground truth labels $(a_k, v_k^a, v_k^c)$, where $a_k$ and $v_k^a, v_k^c$ are labels for answer span prediction and dialog act prediction respectively.

\subsection{Model Overview}
\label{subsec:model-overview}
In the following sections, we present our model that tackles the two tasks described in Section~\ref{subsec:task} together. A summary of key notations is presented in Table~\ref{tab:notations}.

\begin{table}[ht]
\footnotesize
\caption{A summary of key notations used in this paper.}
\label{tab:notations}
\vspace{-0.4cm}
\begin{tabular}{@{}l|l@{}}
\toprule
Notation & Description \\ \midrule
$q_{k}$, $p$ & The $k$-th (current) question in a dialog and the given passage \\
$\mathbf{H}_k$, $\mathbf{H}_k^i$ & The conversation history for $q_k$ and the $i$-th history turn \\
$a_k$, $a_i$ & The ground truth answer for $q_k$ and a history answer for $q_i$ \\
$v_k^a$, $v_k^c$ & The ground truth affirmation and continuation dialog acts for $q_k$ \\
$|V_a|$, $|V_c|$ & The number of classes for affirmation and continuation dialog acts \\
$n$ & The number of ``sub-passages'' after applying a sliding window to $p$ \\
$V_{PosHAE}$ & The vocabulary for PosHAE \\
$\mathbf{ET}$ & The embedding look up table for PosHAE\\
$h$ & The hidden size for PosHAE, $\mathbf{B}$, $\mathbf{E}$, and $\mathbf{D}$ \\
$\mathbf{T}_k^i$, $\mathcal{T}_k$  & One and a batch of contextualized token-level representation(s) \\
$\mathbf{s}_k^i$, $\mathbf{S}_k$,  & One and a batch of contextualized sequence-level representation(s) \\
$I$ & The max \# history turns, which is the first dimension for $\mathcal{T}_k$ and $\mathbf{S}_k$ \\
$F(\cdot)$ & The encoder is a transformation function that $\mathbf{T}_k^i, \mathbf{s}_k^i$ = $F(q_{k}, p, \mathbf{H}_k^i)$ \\
$\mathbf{D}$ & The attention vector in the history attention module\\
$\mathbf{w}, w_i$ & History attention weights and one of the weights \\
$\hat{\mathbf{T}}_k$, $\hat{\mathbf{s}}_k$ & Aggregated token- and sequence-level representations for $\mathcal{T}_k$ and $\mathbf{S}_k$ \\
$\mathbf{t}_k^i(m)$ & The token representation for the $m$-th token in $\mathbf{T}_k^i$ \\
$\mathbf{t}_k(m)$ & All token representations in $\mathcal{T}_k$ for the $m$-th token \\
$\hat{\mathbf{t}}_k(m)$ & The aggregated token rep computed by applying $\mathbf{w}$ to $\{\mathbf{t}_k^i(m)\}_{i=1}^I$ \\
$M$ & The sequence length, which means $\mathbf{T}_k^i$ consists of $M$ tokens \\
$\mathbf{B}$, $\mathbf{E}$ & The begin and end vectors in answer span prediction \\
$p_m^B$, $p_m^E$ & The probabilities of the $m$-th token in $\hat{\mathbf{T}}_k$ being the begin/end tokens\\
$\mathcal{L}_B$, $\mathcal{L}_E$ & The begin and end losses\\
$\mathbf{A}$, $\mathbf{C}$ & Parameters for the affirmation and continuation dialog act predictions\\
$\mathcal{L}_A$, $\mathcal{L}_C$ & Losses for two dialog act predictions\\
$\mathcal{L}_{ans}$, $\mathcal{L}$ & The loss for answer span prediction and the total loss \\
$\lambda$, $\mu$ & Factors to combine $\mathcal{L}_{ans}$, $\mathcal{L}_A$, $\mathcal{L}_C$ to generate $\mathcal{L}$\\ \bottomrule
\end{tabular}
\end{table}

Our proposed model consists of four components: an encoder, a history attention module, an answer span predictor, and a dialog act predictor. The encoder is a BERT model that encodes the question $q_{k}$, the passage $p$, and conversation histories $\mathbf{H}_k$ into contextualized representations. Then the history attention module learns to attend to history turns with different weights and computes aggregated representations for $(q_{k}, p, \mathbf{H}_k)$ on a token level and a sequence level. Finally, the two prediction modules make predictions based on the aggregated representations with a multi-task learning setting. 

\begin{figure*}[t]
    \centering
    \includegraphics[width=0.99\textwidth]{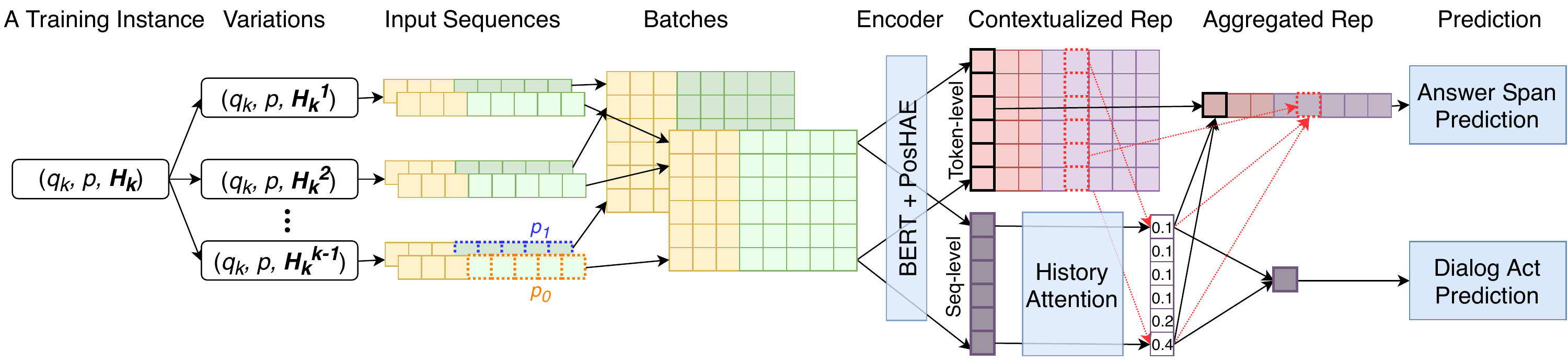}
    \vspace{-0.3cm}
    \caption{Our model consists of an encoder, a history attention module, an answer span predictor, and a dialog act predictor. Given a training instance, we first generate variations of this instance, where each variation contains the same question and passage, with only one turn of conversation history. We use a sliding window approach to split a long passage into ``sub-passages'' ($p_0$ and $p_1$) and use $p_0$ for illustration. The BERT encoder encodes the variations to contextualized representations on both token level and sequence level. The sequence-level representations are used to compute history attention weights. Alternatively, we propose a fine-grained history attention approach as marked in red-dotted lines. Finally, answer span prediction and dialog act predictions are conducted on the aggregated representations generated by the history attention module.}
    \label{fig:model}
    
\end{figure*}

In our architecture, history modeling is enabled in the BERT encoder, where we model one history turn at a time. History selection is performed in the history attention module in the form of ``soft selection''. Figure~\ref{fig:model} gives an overview of our model. We illustrate each component in detail in the following sections.

\subsection{Encoder}
\label{subsubsec:encoder}
\subsubsection{\textbf{BERT Encoder}}
\label{subsubsec:bert-encoder}

\begin{figure}[ht]
    \centering
    \includegraphics[width=0.47\textwidth]{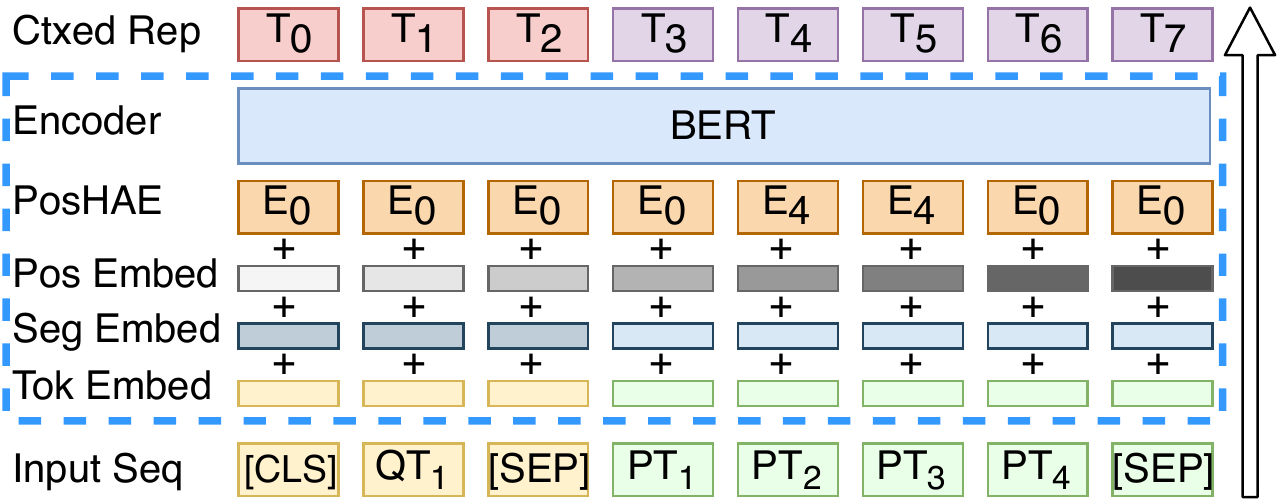}
    \vspace{-0.3cm}
    \caption{The encoder with PosHAE. It zooms in to the encoder in Fig.~\ref{fig:model}. It reveals the encoding process (marked by the blue-dotted lines) from an input sequence (the yellow-green row to the left of the encoder in Fig.~\ref{fig:model}) to contextualized representations (the pink-purple row to the right of the encoder in Fig.~\ref{fig:model}). QT$_i$/PT$_i$ denote question/passage tokens. 
    Suppose we are encoding $(q_{6}, p, \mathbf{H}_6^2)$, $E_4$ and $E_0$ are the history embeddings for tokens that are in and not in $\mathbf{H}_6^2)$.
    }
    \label{fig:encoder}
\end{figure}

The encoder is a BERT model that encodes the question $q_{k}$, the passage $p$, and conversation histories $\mathbf{H}_k$ into contextualized representations. BERT is a pre-trained language model that is designed to learn deep bidirectional representations using transformers~\cite{transformer}. Figure~\ref{fig:encoder} gives an illustration of the encoder. It zooms in to the encoder component in Figure~\ref{fig:model}. It reveals the encoding process from an input sequence (the yellow-green row to the left of the encoder in Figure~\ref{fig:model}) to a contextualized representation (the pink-purple row to the right of the encoder in Figure~\ref{fig:model}).

Given a training instance $(q_{k}, p, \mathbf{H}_k)$, we first generate $k-1$ variations of this instance, where each variation contains the same question and passage, with only one turn of conversation history. Formally, the $i$-th variation is denoted as $(q_{k}, p, \mathbf{H}_k^i)$, where $\mathbf{H}_k^i = (q_i, a_i)$. We follow the previous work~\cite{bert} and use a sliding window approach to split long passages, and thus construct multiple input sequences for a given instance variation. Suppose the passage is split into $n$ pieces,\footnote{$n=2$ in Figure~\ref{fig:model}} the training instance $(q_{k}, p, \mathbf{H}_k)$ would generate $n(k-1)$ input sequences. We take the $k-1$ input sequences corresponding to the first piece of the passage (still denoted as $p$ here for simplicity) for illustration here. As shown in Figure~\ref{fig:encoder}, we pack the question $q_k$ and the passage $p$ into one sequence. The input sequences are fed into BERT and BERT generates contextualized token-level representations for each sequence based on the embeddings for tokens, segments, positions, and a special positional history answer embedding (PosHAE). PosHAE embeds the history answer $a_i$ into the passage $p$ since $a_i$ is essentially a span of $p$. This technique enhances the previous work~\cite{hae} by integrating history position signals. We describe this method in the next section. 

The encoder can be formulated as a transformation function $F(\cdot)$ that takes in a training instance variation and produces a hidden representation for it on a token level, i.e., $\mathbf{T}_k^i$ = $F(q_{k}, p, \mathbf{H}_k^i)$, where $\mathbf{T}_k^i \in \mathbb{R}^{M \times h}$ is the token-level representation for this instance variation. $M$ is the sequence length, and $h$ is the hidden size of the token representation. $\mathbf{T}_k^i$ can also be represented as $\{\mathbf{t}_k^i(m)\}_{m=1}^M$, where $\mathbf{t}_k^i(m) \in \mathbb{R}^{h}$ refers to the representation of the $m$-th token in $\mathbf{T}_k^i$. Instead of using separate encoders for questions, passages, and histories in previous work~\cite{sdnet,flowqa}, we take advantage of BERT and PosHAE to model these different input types jointly.

In addition, we also obtain a sequence-level representation $\mathbf{s}_k^i \in \mathbb{R}^{h}$ for each sequence. We take the representation of the \texttt{[CLS]} token, which is the first token of the sequence, and pass it through a fully-connected layer that has $h$ hidden units~\cite{bert}. That is, $\mathbf{s}_k^i = \text{tanh}(\mathbf{t}_k^i(1) \cdot \mathbf{W}_\text{CLS})$, where $\mathbf{W}_\text{CLS} \in \mathbb{R}^{h \times h}$ is the weight matrix for this dense layer. The bias term in this equation and following equations are omitted for simplicity. This is a standard technique to obtain a sequence-level representation in BERT. It is essentially a pooling method to remove the dimension of sequence length. We also conduct experiments with average pooling and max pooling on this dimension to achieve the same purpose.

\subsubsection{\textbf{Positional History Answer Embedding}}
\label{subsubsec:poshae}
One of the key functions of the encoder is to model the given history turn along with the question and the passage. Previous work~\cite{hae} introduces a history answer embedding (HAE) method to incorporate the conversation history into BERT in a natural way. They learn two unique history answer embeddings that denote whether a token is part of history answers or not. This method gives tokens extra embedding information and thus impacts the token-level contextual representations generated by BERT. However, this method fails to consider the position of a history utterance in the dialog. A commonly used history selection method is to select immediate previous turns. The intuition is that the utility of a history utterance could be related to its position. Therefore, we propose to consider the position information in HAE, resulting in a \textit{positional history answer embedding} (PosHAE) method. The ``position'' refers to the relative position of a history turn in terms of the current question. Our method only considers history answers since previous works~\cite{quac,hae} show that history questions contribute little to the performance. 

Specifically, we first define a vocabulary of size $I + 1$ for PosHAE, denoted as $V_{PosHAE} = \{0, 1, \dots, I\}$, where $I$ is the max number of history turns.\footnote{In QuAC, $I=11$, which means a dialog has at most 11 history turns.} Given the current question $q_k$ and a history turn $H_k^{i}$, we compute the relative position of $H_k^{i}$ in terms of $q_k$ as $k - i$. This relative position corresponds to a vocabulary ID in $V_{PosHAE}$. We use the vocabulary ID $0$ for the tokens that are not in the given history. We then use a truncated normal distribution to initialize an embedding look up table $\mathbf{ET} \in \mathbb{R}^{|I+1| \times h}$. We use $V_{PosHAE}$ to map each token to a history answer embedding in $\mathbf{ET}$. The history answer embeddings are learned. An example is illustrated in Figure~\ref{fig:encoder}. In addition to introducing conversation history, PosHAE enhances HAE by incorporating position information of history turns. This enables the ConvQA model to capture the spatial patterns of history answers in context.

\subsection{\textbf{History Attention Module}}
\label{subsec:history-attention-network}
The core of the history attention module is a  history attention mechanism (HAM). The inputs of this module are the token-level and sequence-level representations for all variations that are generated by the same training instance. The token-level representation is denoted as $\mathcal{T}_k = \{\mathbf{T}_k^i\}_{i=1}^I$, where $\mathcal{T}_k \in \mathbb{R}^{I \times M \times h}$. Similarly, the sequence-level representation is denoted as $\mathbf{S}_k = \{\mathbf{s}_k^i\}_{i=1}^I$, where $\mathbf{S}_k \in \mathbb{R}^{I \times h}$. 
The first dimension of $\mathcal{T}_k$ and $\mathbf{S}_k$ are both $I$ because they are always padded to the max number of history turns. The padded parts are masked out. $\mathcal{T}_k$ and $\mathbf{S}_k$ are illustrated in Figure~\ref{fig:model} as the ``Token-level'' and ``Seq-level Contextualized Rep'' respectively. 

The history attention network is a single-layer feed-forward network. We learn an attention vector $\mathbf{D} \in \mathbb{R}^{h}$ to map a sentence representation $\mathbf{s}_k^i$ to a logit and use the softmax function to compute probabilities across all sequences generated by the same instance. Formally, the history attention weights are computed as follows.
\begin{equation}\label{eqn:attention-weights}
\footnotesize
\begin{aligned}
& w_i = \frac{e^{\mathbf{D} \cdot \mathbf{s}_k^i}}{\sum_{i'=1}^I e^{\mathbf{D} \cdot \mathbf{s}_k^{i'}}} \\
\end{aligned}
\end{equation}
where $w_i$ is the history attention weight for $\mathbf{s}_k^i$. Let $\mathbf{w} = \{w_i\}_{i=1}^I$. We compute aggregated representations for $\mathcal{T}_k$ and $\mathbf{S}_k$ with $\mathbf{w}$:
\begin{equation}\label{eqn:weighted-sum}
\footnotesize
\begin{aligned}
& \hat{\mathbf{T}}_k = \sum_{i=1}^I \mathbf{T}_k^i \cdot w_i \quad , \quad
& \hat{\mathbf{s}}_k = \sum_{i=1}^I \mathbf{s}_k^i \cdot w_i
\end{aligned}
\end{equation}
where $\hat{\mathbf{T}}_k \in \mathbb{R}^{M \times h}$ and $\hat{\mathbf{s}}_k \in \mathbb{R}^{h}$ are aggregated token-level and sequence-level representations respectively. The attention weights $\{w_i\}_{i=1}^I$ are computed on a sequence-level and thus the tokens in the same sequence share the same weight. Intuitively, the history attention network attends to the variation representations with different weights and then each variation representation contributes to the aggregated representation according to the utility of the history turn in this variation.

Alternatively, we develop a \textit{fine-grained history attention} approach to compute the attention weights. Instead of using sequence-level representations $\mathbf{S}_k$ as the input for the attention network, we use the token-level ones. The token-level attention input for the $m$-th token in the sequence is denoted as $\mathbf{t}_k(m) = \{\mathbf{t}_k^i(m)\}_{i=1}^I$, where $\mathbf{t}_k(m) \in \mathbb{R}^{I \times h}$. This is marked as a column with red-dotted lines in Figure~\ref{fig:model}. Then these attention weights are applied to $\mathbf{t}_k(m)$ itself:
\begin{equation}\label{eqn:fine-grained-attention-weights}
\footnotesize
\begin{aligned}
& w_i = \frac{e^{\mathbf{D} \cdot \mathbf{t}_k^i(m)}}{\sum_{i'=1}^I e^{\mathbf{D} \cdot \mathbf{t}_k^{i'}(m)}} \\
& \hat{\mathbf{t}}_k(m) = \sum_{i=1}^I \mathbf{t}_k^i(m) \cdot w_i
\end{aligned}
\end{equation}
where $\hat{\mathbf{t}}_k(m) \in \mathbb{R}^{h}$ is the aggregated token representation for the $m$-th token in this sequence. Therefore, the aggregated token-level representation $\hat{\mathbf{T}}_k$ for this sequence is $\{\hat{\mathbf{t}}_k(m)\}_{m=1}^M$. We show the process of computing the aggregated token representation for one token, but the actual process is vectorized and paralleled for all tokens in this sequence. Intuitively, this approach computes the attention weights given different token representations for the same token but embedded with different history information. These attention weights are on a token level and thus are more fine-grained than those from the sequence-level representations.

In both granularity levels of history attention, we show the process of computing attention weights for a single instance, but the actual process is vectorized for multiple instances. Also, if the given question does not have history turns (i.e., the first question of a conversation), it should bypass the history attention module. In practice, this is equivalent to pass it though the history attention network since all the attention weights will be applied to itself.

\subsection{Answer Span Prediction}
\label{subsec:answer-span-prediction}
Given the aggregated token-level representation $\hat{\mathbf{T}}_k$ produced by the history attention network, we predict answer span by computing the probability of each token being the begin token and the end token. Specifically, we learn two sets of parameters, a begin vector and an end vector, to map a token representation to a logit. Then we use the softmax function to compute probabilities across all tokens in this sequence. Formally, let $\mathbf{B} \in \mathbb{R}^h$ and $ \mathbf{E} \in \mathbb{R}^h$ be the begin vector and the end vector respectively. The probabilities of this token being the begin token $p_m^B$ and end token $p_m^E$ are: 
\begin{equation}\label{eqn:start-prob}
\footnotesize
\begin{aligned}
& p_m^B = \frac{e^{\mathbf{B} \cdot \hat{\mathbf{t}}_k(m)}}{\sum_{m'=1}^M e^{\mathbf{B} \cdot \hat{\mathbf{t}}_k(m')}} \quad , \quad
& p_m^E = \frac{e^{\mathbf{E} \cdot \hat{\mathbf{t}}_k(m)}}{\sum_{m'=1}^M e^{\mathbf{E} \cdot \hat{\mathbf{t}}_k(m')}}
\end{aligned}
\end{equation}
We then compute the cross-entropy loss for answer span prediction:
\begin{equation}\label{eqn:convqa-loss}
\footnotesize
\begin{aligned}
\mathcal{L}_B = - \sum_{M} \mathds{1}\{m=m_B\} &\log{p_m^B} 
\quad , \quad
\mathcal{L}_E = - \sum_{M} \mathds{1}\{m=m_E\} \log{p_m^E} \\
& \mathcal{L}_{ans} = \frac{1}{2} (\mathcal{L}_B + \mathcal{L}_E)
\end{aligned}
\end{equation}
where tokens at positions of $m_B$ and $m_E$ are the ground truth begin token and end token respectively, and  $\mathds{1}\{\cdot\}$ is an indicator function. $\mathcal{L}_B$ and $\mathcal{L}_E$ are the losses for the begin token and end token respectively and $\mathcal{L}_{ans}$ is the loss for answer span prediction. For unanswerable questions, a ``CANNOTANSWER'' token is appended to each passage in QuAC. The model learns to predict an answer span of this exact token if it believes the question is unanswerable.

Invalid predictions, including the cases where the predicted span overlaps with the question part of the sequence, or the end token comes before the begin token, are discarded at testing time.

\subsection{Dialog Act Prediction}
\label{subsec:dialog-act-prediction}
Given the aggregated sequence-level representation $\hat{\mathbf{s}}_k$ for a training instance, we learn two sets of parameters $\mathbf{A} \in \mathbb{R}^{|V_a| \times h}$ and $\mathbf{C} \in \mathbb{R}^{|V_c| \times h}$ to predict the dialog act of affirmation and continuation respectively, where $|V_a|$ and $|V_c|$ denote the number of classes.\footnote{$|V_a|=3$ and $|V_c| =3$ in QuAC.}
Formally, the loss for dialog act prediction for affirmation is:
\begin{equation}\label{eqn:dialog-act-loss}
\footnotesize
\begin{aligned}
& p(v|\hat{\mathbf{s}}_k) = \frac{e^{\mathbf{A}_v \cdot \hat{\mathbf{s}}_k}}{\sum_{v'=1}^{|V_a|} e^{\mathbf{A}_{v'} \cdot \hat{\mathbf{s}}_k}} \\
& \mathcal{L}_A = - \sum_{v} \mathds{1}\{v=v_k^a\} \log{p(v|\hat{\mathbf{s}}_k)}
\end{aligned}
\end{equation}
where $\mathds{1}\{\cdot\}$ is an indicator function to show whether the predicted label $v$ is the ground truth label $v_k^a$, and $\mathbf{A}_v \in \mathbb{R}^{h}$ is the vector in $\mathbf{A}$ corresponding to $v$. The loss $\mathcal{L}_C$ for predicting the continuation dialog act $v_k^c$ is computed in the same way. We make dialog act predictions independently based on the information of each single training instance $(q_{k}, p, \mathbf{H}_k)$. We do not model history dialog acts in the encoder for this task.

\subsection{Model Training}
\label{subsec:model-training}
\subsubsection{\textbf{Batching}}
We implement an \textit{instance-aware batching} approach to construct the batches for BERT. This method guarantees that the variations generated by the same training instance are always included in the same batch, so that the history attention module operates on all available histories. In practice,
a passage in a training instance can produce multiple ``sub-passages'' (e.g., $p_0$ and $p_1$ in Figure~\ref{fig:model}) after applying the sliding window approach~\cite{bert}. This results in multiple ``sub-instances'' (e.g. $(q_{k}, p_0, \mathbf{H}_k^i)$ and $(q_{k}, p_1, \mathbf{H}_k^i)$), which are modeled separately and potentially in different batches. This is because the ``sub-passages'' have overlaps to make sure that every passage token has sufficient context so that they can be considered as different passages.

\subsubsection{\textbf{Training Loss and Multi-task Learning}}
We adopt the multi-task learning idea to jointly learn the answer span prediction task and the dialog act prediction task. All parameters are learned in an end-to-end manner. We use hyper-parameters $\lambda$ and $\mu$ to combine the losses for different tasks. That is,
\begin{equation}\label{eqn:mtl-loss}
\begin{aligned}
\mathcal{L} = \mu \mathcal{L}_{ans} + \lambda \mathcal{L}_A + \lambda \mathcal{L}_C
\end{aligned}
\end{equation}
where $\mathcal{L}$ is the total training loss. 

Multi-task learning has been shown to be effective for representation learning~\cite{Liu2019MultiTaskDN,Liu2015RepresentationLU,Xu2018MultiTaskLF}. There are two reasons behind this. 1) Our two tasks provide more supervising signals to fine-tune the encoder. 2) Representation learning benefits from a regularization effect by optimizing for multiple tasks. Although BERT serves as a universal encoder by pre-training with a large amount of unlabeled data, MTL is a complementing technology~\cite{Liu2019MultiTaskDN} that makes such representations more generic and transferable. More importantly, we can handle two essential tasks in ConvQA, answer span prediction and dialog act prediction, with a uniform model architecture.

%% file: 4_experiments.tex
\section{Experiments}
\label{sec:exp}

\subsection{Data Description}
\label{subsec:data}

We experiment with the QuAC (Question Answering in Context) dataset~\cite{quac}. It is a large-scale dataset designed for modeling and understanding information-seeking conversations. It contains interactive dialogs between an information-seeker and an information-provider. The information-seeker tries to learn about a \textit{hidden} Wikipedia passage by asking a sequence of freeform questions. She/he only has access to the heading of the passage, simulating an information need. The information-provider answers each question by providing a short span of the given passage. One of the unique properties that distinguish QuAC from other dialog data is that it comes with dialog acts. The information-provider uses dialog acts to provide the seeker with feedback (e.g., ``ask a follow up question''), which makes the dialogs more productive~\cite{quac}. This dataset poses unique challenges because its questions are more open-ended, unanswerable, or only meaningful within the dialog context. More importantly, many questions have coreferences and interactions with conversation history, making this dataset suitable for our task. We present some statistics of the dataset in Table~\ref{tab:data-stat}.
\begin{table}[htbp]
\footnotesize
\caption{Data Statistics. We can only access the training and validation data.}
\label{tab:data-stat}
\vspace{-0.4cm}
\begin{tabular}{@{}lll@{}}
\toprule
Items                                            & Train      & Validation \\ \midrule
\# Dialogs                                       & 11,567     & 1,000      \\
\# Questions                                     & 83,568     & 7,354      \\
\# Average Tokens Per Passage                    & 396.8      & 440.0      \\
\# Average Tokens Per Question                   & 6.5        & 6.5        \\
\# Average Tokens Per Answer                     & 15.1       & 12.3       \\
\# Average Questions Per Dialog                  & 7.2        & 7.4        \\
\# Min/Avg/Med/Max History Turns Per Question & 0/3.4/3/11 & 0/3.5/3/11 \\ \bottomrule
\end{tabular}
\end{table}

\subsection{Experimental Setup}
\label{subsec:setup}
\subsubsection{\textbf{Competing Methods}}
\label{subsubsec:competing-methods}
We consider all methods with published papers on the QuAC leaderboard as baselines.\footnote{The methods without published papers or descriptions are essentially done in parallel with ours and may not be suitable for comparison since their model details are unknown. Besides, these work could be using generic performance boosters, such as BERT-large, data augmentation, transfer learning, or better training infrastructures.} In addition, we also include a ``BERT + PosHAE'' model that replaces HAE in \citet{hae} with PosHAE to demonstrate the impact of the PosHAE. To be specific, the competing methods are:
\begin{itemize}[leftmargin=1em]
    \item \textbf{BiDAF++}~\cite{Peters2018DeepCW,quac}: BiDAF~\cite{bidaf} is a top-performing SQuAD model. It uses bi-directional attention flow mechanism to obtain a query-aware context representation. BiDAF++ makes further augmentations with self-attention~\cite{Clark2018SimpleAE} and contextualized embeddings. 
    
    \item \textbf{BiDAF++ w/ 2-Context}~\cite{quac}: This model incorporates conversation history by modifying the passage and question embedding processes. Specifically, it encodes the dialog turn number with the question embedding and concatenates answer marker embeddings to the word embedding. 
    
    \item \textbf{FlowQA}~\cite{flowqa}: This model incorporates conversation history by integrating intermediate representation generated when answering the previous question. Thus it is able to grasp the latent semantics of the conversation history compared to shallow approaches that concatenate history turns.
    
    \item \textbf{BERT + HAE}~\cite{hae}: This model is adapted from the SQuAD model in the BERT paper.\footnote{We notice the hyper-parameter of ``max answer length'' is set to 30 in BERT + HAE~\cite{hae}, which is sub-optimal. We set it to 40 to be consistent with our settings and updated their validation results.} It uses history answer embedding to enable a seamless integration of conversation history into BERT.
    
    \item \textbf{BERT + PosHAE}: We enhance the BERT + HAE model with the PosHAE that we proposed. This method considers the position information of history turns and serves as a stronger baseline. We set the max number of history turns as 6 since it gives the best performance under this setting.
    
    \item \textbf{HAM} (History Attention Mechanism): This is the solution we proposed in Section~\ref{sec:our-approach}. It employs PosHAE for history modeling, the history attention mechanism for history selection, and the MTL scheme to optimize for both answer span prediction and dialog act prediction tasks. We use the fine-grained history attention in Equation~\ref{eqn:fine-grained-attention-weights}. We use ``HAM'' as the model name since the attentive history selection is the most important and effective component that essentially defines the model architecture.
    
    \item \textbf{HAM (BERT-Large)}: Due to the competing nature of the QuAC challenge, we apply BERT-Large to HAM for a more informative evaluation. This is more resource intensive. Other HAM models in this paper are constructed with BERT-Base for two reasons: 1) To alleviate the memory and training efficiency issues caused by BERT-Large and thus speed up the experiments for the research purpose. 2) To keep the settings consistent with existing and published work~\cite{hae} for fair and easy comparison. 

\end{itemize}

\subsubsection{\textbf{Evaluation Metrics}}
\label{subsubsec:metrics}
The QuAC challenge provides two evaluation metrics, the word-level F1, and the human equivalence score (HEQ)~\cite{quac}. The word-level F1 evaluates the overlap of the prediction and the ground truth answer span. It is a classic metric used in MC and ConvQA tasks~\cite{squad,coqa,quac}. HEQ measures the percentage of examples for which system F1 exceeds or matches human F1. Intuitively, this metric judges whether a system can provide answers as good as an average human. This metric is computed on the question level (HEQ-Q) and the dialog level (HEQ-D). In addition, the dialog act prediction task is evaluated by accuracy.

\subsubsection{\textbf{Hyper-parameter Settings and Implementation Details}}
\label{subsubsec:details}
Models are implemented with TensorFlow\footnote{\url{https://www.tensorflow.org/}}. The version of the QuAC data we use is v0.2. We use the BERT-Base Uncased model\footnote{\url{https://github.com/google-research/bert}} with the max sequence length set to 384. The batch size is set to 24. We train the ConvQA model with a Adam weight decay optimizer with an initial learning rate of 3e-5. The warming up portion for learning rate is 10\%. We set the stride in the sliding window for passages to 128, the max question length to 64, and the max answer length to 40. The total training steps is set to 30,000. Experiments are conducted on a single NVIDIA TESLA M40 GPU. $\lambda$ and $\mu$ for multi-task learning is set to 0.1 and 0.8 respectively for HAM. 

\subsection{Main Evaluation Results}
\label{subsec:results}
We report the results on the validation and test sets in Table~\ref{tab:results}. Our best model was evaluated officially by the QuAC challenge and the result is displayed on the leaderboard\footnote{\url{http://quac.ai/}} with proper anonymization. Since dialog act prediction is not the main task of this dataset, most of the baseline methods do not perform this task.

\begin{table}[htbp]
\caption{Evaluation results on QuAC. Models in a bold font are our implementations. Each cell displays val/test scores. Val result of BiDAF++, FlowQA are from \cite{quac}, \cite{flowqa}. Test results are from the QuAC leaderboard at the time of the CIKM deadline. $\ddagger$ means statistically significant improvement over the strongest baseline with $p < 0.05$ tested by the Student's paired t-test. We can only do significance test on F1 on the validation set. ``--'' means a result is not available and ``N/A'' means a result is not applicable for this model.
}
\label{tab:results}
\vspace{-0.3cm}
\footnotesize
\tabcolsep=0.08cm
\begin{tabular}{@{}lllllll@{}}
\toprule
Models            & F1           & HEQ-Q        & HEQ-D          & Yes/No & Follow up \\ \midrule
BiDAF++           & 51.8 / 50.2  & 45.3 / 43.3  & 2.0 / 2.2 & 86.4 / 85.4 & 59.7 / 59.0  \\
BiDAF++ w/ 2-C    & 60.6 / 60.1  & 55.7 / 54.8  & 5.3 / 4.0 & 86.6 / 85.7 & 61.6 / 61.3  \\

BERT + HAE        & 63.9 / 62.4  & 59.7 / 57.8  & 5.9 / 5.1 & N/A & N/A  \\ 
FlowQA            & 64.6 / 64.1  & \hspace{0.15cm}--\hspace{0.15cm}   / 59.6  & \hspace{0.1cm}--\hspace{0.1cm} / 5.8 & N/A & N/A \\ \hline
\textbf{BERT + PosHAE }         & 64.7 / \hspace{0.15cm}--\hspace{0.15cm}  & 60.7 / \hspace{0.15cm}--\hspace{0.15cm}  & 6.0 / \hspace{0.15cm}--\hspace{0.15cm} & N/A & N/A  \\ 
\textbf{HAM}     & 65.7$^\ddagger$ / 64.4  & 62.1 / 60.2  & 7.3 / 6.1 & \textbf{88.3} / \textbf{88.4} & 62.3 / \textbf{61.7}  \\
\textbf{HAM (BERT-Large)}     & \textbf{66.7}$^\ddagger$ / \textbf{65.4}  & \textbf{63.3} / \textbf{61.8}  & \textbf{9.5} / \textbf{6.7} & 88.2 / 88.2 & \textbf{62.4} / 61.0  \\
\bottomrule
\end{tabular}
\end{table}

We summarize our observations of the results as follows. 
\begin{enumerate}[leftmargin=1em]
\item BERT + PosHAE brings a significant improvement compared with BERT + HAE, achieving the best results among baselines. This suggests that the position information plays an important role in conversation history modeling with history answer embedding. In addition, previous work reported that BERT + HAE enjoys a much better training efficiency compared to FlowQA but suffers from a poorer performance. However, after enhancing HAE with the history position information, it manages to achieve a slightly higher performance than FlowQA when maintaining the efficiency advantage. This shows the effectiveness of this conceptually simple idea of modeling conversation history in BERT with PosHAE.
\item Our model HAM obtains statistically significant improvements over the strongest baseline (BERT + PosHAE) with $p < 0.05$ tested by the Student's paired t-test. These results demonstrate the effectiveness of our method. 
\item Our model HAM also achieves a substantially higher performance on dialog act prediction compared to baseline methods, showing the strength of our model on both tasks. We can only do significance test on F1. We are unable to do a significance test on dialog act prediction because the prediction results of BiDAF++ is not available. In addition, the sequence-level representations of HAM are obtained with max pooling. We see no major differences when using different pooling methods.
\item Applying BERT-Large to HAM brings a substantial improvement to answer span prediction, suggesting that a more powerful encoder can boost the performance. 
\end{enumerate}

\begin{figure*}[ht]
    \centering
    \begin{subfigure}[b]{0.32\textwidth}
        \includegraphics[width=\textwidth]{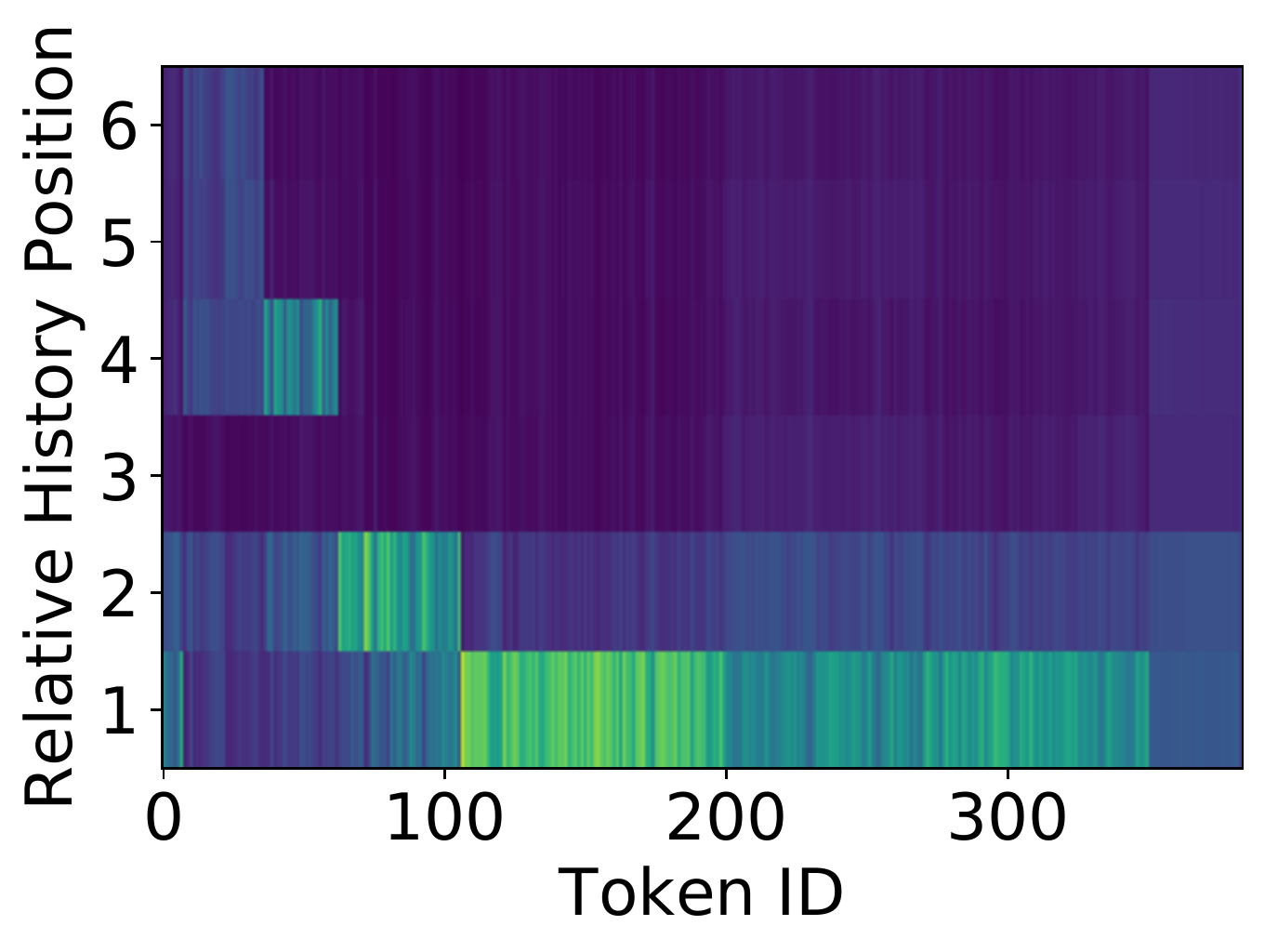}
        \vspace{-0.6cm}
        \caption{Drill down}
        \label{fig:drill-down}
    \end{subfigure}
    \begin{subfigure}[b]{0.32\textwidth}
        \includegraphics[width=\textwidth]{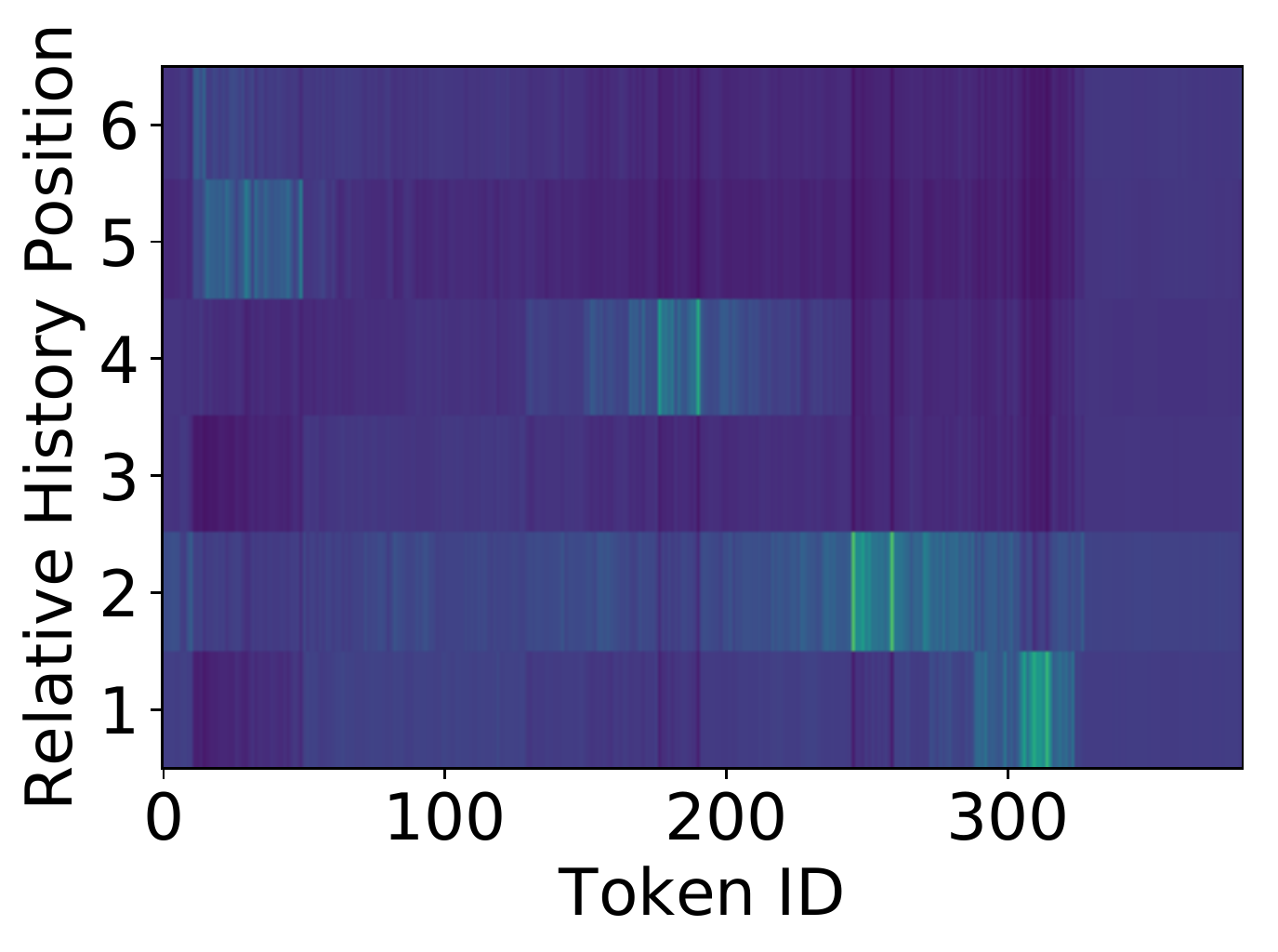}
        \vspace{-0.6cm}
        \caption{Topic shift}
        \label{fig:topic-shift}
    \end{subfigure}
    \begin{subfigure}[b]{0.32\textwidth}
        \includegraphics[width=\textwidth]{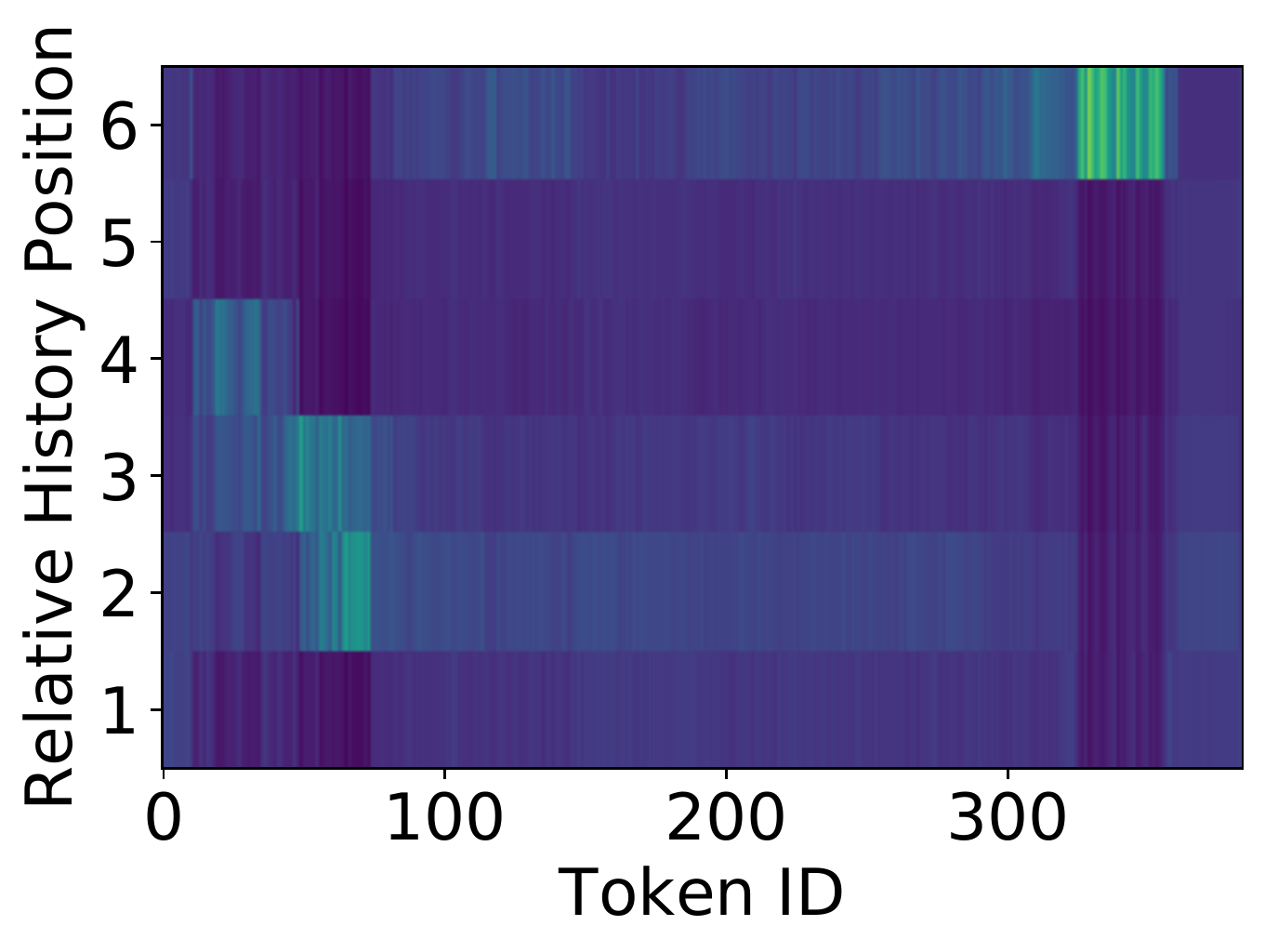}
        \vspace{-0.6cm}
        \caption{Topic return}
        \label{fig:topic-return}
    \end{subfigure}
    \vspace{-0.4cm}
    \caption{Attention visualization for different dialog behaviors. Brighter spots mean higher attention weights. Token ID refers to the token position in an input sequence. A sequence contains 384 tokens. Relative history position refers to the difference of the current turn \# with a history turn \#. The selected examples are all in the 7th turn. These figures are best viewed in color. }
    \label{fig:attention-visualization}
\end{figure*}

\begin{table*}[h]
\tabcolsep=0.08cm
\footnotesize
\caption{QuAC dialogs that correspond to the dialog behaviors in Fig.~\ref{fig:attention-visualization}. The examples are all in the 7th turn. ``\#'' refers to the relative history position, which means ``0'' is the current turn and ``6'' is the most remote turn from the current turn. Each turn has a question and an answer, with the answer in italic. Co-references and related terms are marked in the same color.}\label{tab:dialog-behaviors}
\vspace{-0.4cm}
\begin{subtable}{.33\linewidth}\centering
\caption{Drill down}\label{tab:drill-down}
{

\begin{tabular}{@{}l|l@{}}
\toprule
\#                 & Utterance                                                                         \\ \midrule
\multirow{2}{*}{6} & When did Ride leave NASA?                                                         \\
                   & \textit{In 1987, Ride left ... to work at the Stanford ...}          \\ \midrule
\multirow{2}{*}{5} & What did she do at the Stanford Center?                                           \\
                   & \textit{International Security and Arms Control. }                                \\ \midrule
\multirow{2}{*}{4} & How long was she there?                                                           \\
                   & \textit{In 1989, she became a professor of physics at ...}                        \\ \midrule
\multirow{2}{*}{3} & Was she successful as a professor?                                                \\
                   & \textit{CANNOTANSWER}                                                             \\ \midrule
\multirow{2}{*}{2} & Did she have any other professions?                                               \\
                   & \textit{Ride led two public-outreach \textcolor{blue}{\underline{programs}} for NASA ...}     \\ \midrule
\multirow{2}{*}{1} & What was involved in the \textcolor{blue}{\underline{programs}}?                              \\
                   & \textit{The \textcolor{blue}{\underline{programs}} allowed middle school students to ...}     \\ \midrule
\multirow{2}{*}{0} & What did she do after \textcolor{blue}{\underline{this}}?                                     \\
                   & \textit{To be predicted ... }                                                     \\ \bottomrule
\end{tabular}
}
\end{subtable}%
\begin{subtable}{.33\linewidth}\centering
\caption{Topic shift}\label{tab:topic-shift}
{
\begin{tabular}{@{}l|l@{}}
\toprule
\#                 & Utterance                                              \\ \midrule
\multirow{2}{*}{6} & When did the \textcolor{violet}{\underline{Greatest Hits}} come out              \\
                   & \textit{beginning of 2004}          \\ \midrule
\multirow{2}{*}{5} & What songs were on the \textcolor{violet}{\underline{album}}                                      \\
                   & \textit{cover of Nick Kamen's ``I Promised Myself'' ...}                                \\ \midrule
\multirow{2}{*}{4} & Was the \textcolor{violet}{\underline{album}} popular                                                        \\
                   & \textit{The single became another top-two hit for the band ...}                        \\ \midrule
\multirow{2}{*}{3} & Did \textcolor{violet}{\underline{it}} win any awards                                      \\
                   & \textit{CANNOTANSWER}                                                             \\ \midrule
\multirow{2}{*}{2} & Why did they release \textcolor{violet}{\underline{this}}                                             \\
                   & \textit{... was just released in selected European countries ...}    \\ \midrule
\multirow{2}{*}{1} & Did they tour with this \textcolor{violet}{\underline{album}}?      \\
                   & \textit{the band finished their tour}  \\ \midrule
\multirow{3}{*}{0} & \textcolor{blue}{Are there other interesting aspects about this article?}\\
                   & \textit{To be predicted ... }                                                     \\ \bottomrule
\end{tabular}
}
\end{subtable}
\begin{subtable}{.33\linewidth}\centering
\caption{Topic return}\label{tab:topic-return}
{
\begin{tabular}{@{}l|l@{}}
\toprule
\#                 & Utterance                                                                         \\ \midrule
\multirow{2}{*}{6} & What is relevant about Lorrie's \textcolor{blue}{\underline{musical career}}?                                                         \\
                   & \textit{... she signed with RCA Records ... her first album ...}          \\ \midrule
\multirow{2}{*}{5} & What songs are included in the album?                                          \\
                   & \textit{CANNOTANSWER}                                \\ \midrule
\multirow{2}{*}{4} & Are there any other interesting aspects about this article?  \\
                   & \textit{made her first appearance on the Grand Ole Opry at age 13,}                        \\ \midrule
\multirow{2}{*}{3} & What did she do after her first appearance?                                               \\
                   & \textit{... she took over ... and began leading the group ...}     \\ \midrule
\multirow{2}{*}{2} & What important work did she do with the band?                                              \\
                   & \textit{leading the group through various club gigs.}     \\ \midrule
\multirow{2}{*}{1} & What songs did she played with the group?                             \\
                   & \textit{CANNOTANSWER}     \\ \midrule
\multirow{2}{*}{0} & What are other interesting aspects of her \textcolor{blue}{\underline{musical career}}?                                     \\
                   & \textit{To be predicted ... }                                                     \\ \bottomrule
\end{tabular}
}
\end{subtable}
    
\end{table*}

\subsection{Ablation Analysis}
\label{subsec:ablation}
Section~\ref{subsec:results} shows the effectiveness of our model. This performance is closely related to several design choices. So we conduct an ablation analysis to investigate the contributions of each design choice by removing or replacing the corresponding component in the complete HAM model. Specifically, we have four settings as follows.
\begin{itemize}[leftmargin=1em]
    \item \textbf{HAM w/o Fine-grained (F-g) History attention}. We use the sequence-level history attention (Equation~\ref{eqn:attention-weights} and \ref{eqn:weighted-sum}) instead of the fine-grained history attention (Equation~\ref{eqn:fine-grained-attention-weights}).
    
    \item \textbf{HAM w/o History Attention}. We do not learn any form of history attention. Instead, we modify the history attention module and make it always produce equal weights. Note that this is not equivalent to ``BERT + PosHAE''. ``BERT + PosHAE'' incorporates the selected history turns in a single input sequence and relies on the encoder to work out the importance of these history turns. The architecture we illustrated in Figure~\ref{fig:model} models each history turn separately and capture their importance by the history attention mechanism explicitly, which is a more direct and explainable way. Therefore, even when we disable the history attention module, it is not equivalent to ``BERT + PosHAE''.
    
    \item \textbf{HAM w/o PosHAE}. We use HAE~\cite{hae} instead of the PosHAE we proposed in Section~\ref{subsubsec:poshae}.
    
    \item \textbf{HAM w/o MTL}. Our multi-task learning scheme consists of two tasks, an answer span prediction task and a dialog act prediction task. Therefore, to evaluate the contribution of MTL, we further design two settings: (1) In \textbf{HAM w/o Dialog Act Prediction}, we set $\mu = 1$ and $\lambda = 0$ in Equation~\ref{eqn:mtl-loss} to block the parameter updates from dialog act prediction. (2) In \textbf{HAM w/o Answer Span Prediction}, we set $\mu = 0$ in Equation~\ref{eqn:mtl-loss} and thus block the updates caused by answer span prediction. We tune $\lambda$ in (0.2, 0.4, 0.6, 0.8) in Equation~\ref{eqn:mtl-loss} and try different pooling methods to obtain the sequence-level representations. We finally adopt $\lambda=0.2$ and average pooling since they give the best performance. We consider these two ablation settings to fully control the factors in our experiments and thus precisely capture the differences in the representation learning caused by different tasks.
\end{itemize}

\begin{table}[htbp]
\caption{Results for ablation analysis. These results are obtained on the validation set since the test set is hidden for official evaluation only. ``w/o'' means to remove or replace the corresponding component. $\dagger$ means statistically significant performance \textit{decrease} compared to the complete HAM model with $p < 0.05$ tested by the Student's paired t-test. We can only do significance test on F1 and dialog act accuracy. }
\label{tab:ablation}
\vspace{-0.4cm}
\footnotesize
\begin{tabular}{@{}lllllll@{}}
\toprule
Models                              & F1    & HEQ-Q  & HEQ-D  & Yes/No  & Follow up \\ \midrule
HAM             & 65.7  & 62.1   & 7.3    & 88.3    & 62.3      \\ \hline
w/o F-g History Attention            & 64.9$^\dagger$  & 61.0   & 7.1    & 88.4    & 62.1      \\
w/o History Attention                & 61.1$^\dagger$  & 57.2    & 6.4     & 87.9       & 60.5$^\dagger$      \\ 
w/o PosHAE                           & 64.2$^\dagger$  & 60.0    & 7.3    & 88.6    & 62.1      \\ 
w/o Dialog Act Prediction            & 65.9  & 62.2    & 8.2    & N/A     & N/A       \\ 
w/o Answer Span Prediction           & N/A   & N/A    & N/A    & 86.2$^\dagger$    & 59.7$^\dagger$       \\ 
\bottomrule
\end{tabular}
\end{table}

The ablation results on the validation set are presented in Table~\ref{tab:ablation}. 
The following are our observations.
\begin{enumerate}[leftmargin=1em, noitemsep]
    \item By replacing the fine-grained history attention with sequence-level history attention, we observe a performance drop. This shows the effectiveness of computing history attention weights on a token level. This is intuitive because these weights are specifically tailored for the given token and thus can better capture the history information embedded in the token representations. 
    
    \item When we disable the history attention module, we notice the performance drops dramatically for 4.6\% and 3.8\% compared with HAM and ``HAM w/o F-g History Attention'' respectively. This indicates that the history attention mechanism, regardless of granularity, can attend to conversation histories according to their importance. Disabling history attention also hurts the performance for dialog act prediction.
    
    \item Replacing PosHAE with HAE also witnesses a major drop in model performance. This again shows the importance of history position information in modeling conversation history.
    
    \item When we remove the dialog act prediction task, we observe that the performance for answer span prediction has a slight and insignificant increase. This suggests that dialog act prediction does not contribute to the representation learning for answer span prediction. Since dialog act prediction is a secondary task in our setting, its loss is scaled down and thus could have a limited impact on the optimization for the encoder. Although the performance for our main model is slightly lower on answer span prediction, it can handle both answer span prediction and dialog prediction tasks in a uniform way. 
    
    \item On the contrary, when we remove the answer span prediction task, we observe a relatively large performance drop for dialog act prediction. This indicates that the additional supervising signals from answer span prediction can indeed help the encoder to produce a more generic representation that benefits the dialog act prediction task. In addition, the encoder could also benefit from a regularization effect because it is optimized for two different tasks and thus alleviates overfitting. Although the multi-task learning scheme does not contribute to answer span prediction, we show that it is beneficial to dialog act prediction. 

\end{enumerate}

\subsection{Case Study and Attention Visualization}
\label{subsec:case-study}
One of the major advantages of our model is its explainability of history attention. In this section, we present a case study that visualizes the history attention weights predicted by our model.

\citet{Qu2018AnalyzingAC} observed that \textit{follow up questions} is one of the most important user intents in information-seeking conversations. \citet{Yatskar2018AQC} further described three history-related dialog behaviors that can be considered as a fine-grained taxonomy of follow up questions. We use these definitions to interpret the attention weights. These dialog behaviors are as follow.
\begin{itemize}[leftmargin=1em]
    \item \textbf{Drill down}: the current question is a request for more information about a topic being discussed.
    \item \textbf{Topic shift}: the current question is not immediately relevant to something previously discussed.
    \item \textbf{Topic return}: the current question is asking about a topic again after it had previously been shifted away from.
\end{itemize}

We keep records of the attention weights generated at testing time on the validation data. We use a sliding window approach to split long passages as mentioned in Section~\ref{subsubsec:bert-encoder}. However, we specifically choose short passages that can be put in a single input sequence for easier visualization. The attention weights obtained from our fine-grained history attention model are visualized in Figure~\ref{fig:attention-visualization} and the corresponding dialogs are presented in Table~\ref{tab:dialog-behaviors}.


Our history attention weights are computed on the token level. We observe that salient tokens are typically in the corresponding history answer in the passage. This suggests that our model learns to attend to tokens that carry history information. These tokens also bring some attention weights to other tokens that are not in the history answer since the token representations are contextualized. Although each history turn has an answer, the weights vary to reflect the importance of the history information. 

We further interpret the attention weights with examples for different dialog behaviors. First, Table~\ref{tab:drill-down} shows that the current question is drilling down on more relevant information on the topic being discussed. In this case, the current question is closely related to its immediate previous turns. We observe in Figure~\ref{fig:drill-down} that our model can attend to these turns properly with greater weights assigned to the most immediate previous turn. Second, in the topic shift scenario presented in Table~\ref{tab:topic-shift} and Figure~\ref{fig:topic-shift}, the current question is not immediately relevant to its preceding history turns. Therefore, the attention weights are distributed relatively evenly across history turns. Third, as shown in Table~\ref{tab:topic-return} and Figure~\ref{fig:topic-return}, the first turn talks about the topic of musical career while the following turns shift away from this topic. The information-seeker returns to musical career in the current turn. In this case, the most important history turn to consider is the most remote one from the current question. Our model learns to attend to certain tokens the first turn with larger weights, suggesting that the model could capture the topic return phenomenon. Moreover, we observe that the model does not attend to the passage token of ``CANNOTANSWER'', further indicating that it can identify useful history answers. 

%% file: 5_conclusions.tex
\section{Conclusions and Future work} 
\label{sec:conclusion}

In this work, we propose a novel model for ConvQA. We introduce a history attention mechanism to conduct a ``soft selection'' for conversation histories. We show that our model can capture the utility of history turns. In addition, we enhance the history answer embedding method by incorporating the position information for history turns. We show that history position information plays an important role in conversation history modeling. Finally, we propose to jointly learn answer span prediction and dialog act prediction with a uniform model architecture in a multi-task learning setting. We conduct extensive experimental evaluations to demonstrate the effectiveness of our model. For future work, we would like to consider to apply our history attention method to other conversational retrieval tasks. 
In addition, 
we will further analyze the relationship between attention patterns and different user intents or dialog acts.

%% file: 0_CQA.bbl
\begin{thebibliography}{39}
\providecommand{\natexlab}[1]{#1}
\providecommand{\url}[1]{\texttt{#1}}
\expandafter\ifx\csname urlstyle\endcsname\relax
  \providecommand{\doi}[1]{doi: #1}\else
  \providecommand{\doi}{doi: \begingroup \urlstyle{rm}\Url}\fi

\bibitem[Belkin et~al.(1994)Belkin, Cool, S., and Thiel]{Belkin1994CasesS}
N.~J. Belkin, C.~Cool, A.~S., and U.~Thiel.
\newblock {Cases , Scripts , and Information-Seeking Strategies : On the Design
  of Interactive Information Retrieval Systems}.
\newblock 1994.

\bibitem[Choi et~al.(2018)Choi, He, Iyyer, Yatskar, Yih, Choi, Liang, and
  Zettlemoyer]{quac}
E.~Choi, H.~He, M.~Iyyer, M.~Yatskar, W.~Yih, Y.~Choi, P.~Liang, and L.~S.
  Zettlemoyer.
\newblock {QuAC: Question Answering in Context}.
\newblock In \emph{EMNLP}, 2018.

\bibitem[Chuklin et~al.(2018)Chuklin, Severyn, Trippas, Alfonseca, Sil{\'e}n,
  and Spina]{Chuklin2018ProsodyMF}
A.~Chuklin, A.~Severyn, J.~R. Trippas, E.~Alfonseca, H.~Sil{\'e}n, and
  D.~Spina.
\newblock {Prosody Modifications for Question-Answering in Voice-Only
  Settings}.
\newblock \emph{CoRR}, 2018.

\bibitem[Clark and Gardner(2018)]{Clark2018SimpleAE}
C.~Clark and M.~Gardner.
\newblock {Simple and Effective Multi-Paragraph Reading Comprehension}.
\newblock In \emph{ACL}, 2018.

\bibitem[Croft and Thompson(1987)]{i3r}
W.~B. Croft and R.~H. Thompson.
\newblock {I3R: A new approach to the design of document retrieval systems}.
\newblock \emph{JASIS}, 38:\penalty0 389--404, 1987.

\bibitem[Devlin et~al.(2018)Devlin, Chang, Lee, and Toutanova]{bert}
J.~Devlin, M.-W. Chang, K.~Lee, and K.~Toutanova.
\newblock {BERT: Pre-training of Deep Bidirectional Transformers for Language
  Understanding}.
\newblock \emph{CoRR}, 2018.

\bibitem[Gao et~al.(2018)Gao, Galley, and Li]{Gao2018NeuralAT}
J.~Gao, M.~Galley, and L.~Li.
\newblock {Neural Approaches to Conversational AI}.
\newblock In \emph{SIGIR}, 2018.

\bibitem[Guo et~al.(2019)Guo, Fan, Pang, Yang, Ai, Zamani, Wu, Croft, and
  Cheng]{Guo2019ADL}
J.~Guo, Y.~Fan, L.~Pang, L.~Yang, Q.~Ai, H.~Zamani, C.~Wu, W.~B. Croft, and
  X.~Cheng.
\newblock {A Deep Look into Neural Ranking Models for Information Retrieval}.
\newblock \emph{CoRR}, abs/1903.06902, 2019.

\bibitem[Hu et~al.(2018)Hu, Peng, Huang, Qiu, Wei, and
  Zhou]{Hu2018ReinforcedMR}
M.~Hu, Y.~Peng, Z.~Huang, X.~Qiu, F.~Wei, and M.~Zhou.
\newblock {Reinforced Mnemonic Reader for Machine Reading Comprehension}.
\newblock In \emph{IJCAI}, 2018.

\bibitem[Huang et~al.(2017)Huang, Zhu, Shen, and Chen]{Huang2017FusionNetFV}
H.-Y. Huang, C.~Zhu, Y.~Shen, and W.~Chen.
\newblock {FusionNet: Fusing via Fully-Aware Attention with Application to
  Machine Comprehension}.
\newblock \emph{CoRR}, abs/1711.07341, 2017.

\bibitem[Huang et~al.(2018)Huang, Choi, and Yih]{flowqa}
H.-Y. Huang, E.~Choi, and W.~Yih.
\newblock {FlowQA: Grasping Flow in History for Conversational Machine
  Comprehension}.
\newblock \emph{CoRR}, 2018.

\bibitem[Joshi et~al.(2017)Joshi, Choi, Weld, and Zettlemoyer]{TriviaQA}
M.~S. Joshi, E.~Choi, D.~S. Weld, and L.~S. Zettlemoyer.
\newblock {TriviaQA: A Large Scale Distantly Supervised Challenge Dataset for
  Reading Comprehension}.
\newblock In \emph{ACL}, 2017.

\bibitem[Kotov and Zhai(2010)]{Kotov2010TowardsNQ}
A.~Kotov and C.~Zhai.
\newblock {Towards natural question guided search}.
\newblock In \emph{WWW}, 2010.

\bibitem[Kwiatkowski et~al.(2019)Kwiatkowski, Palomaki, Redfield, Collins,
  Parikh, Alberti, Epstein, Polosukhin, Kelcey, Devlin, Lee, Toutanova, Jones,
  Chang, Dai, Uszkoreit, Le, and Petrov]{GoogleNQ}
T.~Kwiatkowski, J.~Palomaki, O.~Redfield, M.~Collins, A.~Parikh, C.~Alberti,
  D.~Epstein, I.~Polosukhin, M.~Kelcey, J.~Devlin, K.~Lee, K.~N. Toutanova,
  L.~Jones, M.-W. Chang, A.~Dai, J.~Uszkoreit, Q.~Le, and S.~Petrov.
\newblock {Natural Questions: a Benchmark for Question Answering Research}.
\newblock \emph{Transactions of the Association of Computational Linguistics},
  2019.

\bibitem[Li et~al.(2017)Li, Qiu, Chen, Wang, Gao, Huang, Ren, Zhao, Zhao, Wang,
  Jin, and Chu]{Li2017AliMeA}
F.-L. Li, M.~Qiu, H.~Chen, X.~Wang, X.~Gao, J.~Huang, J.~Ren, Z.~Zhao, W.~Zhao,
  L.~Wang, G.~Jin, and W.~Chu.
\newblock {AliMe Assist : An Intelligent Assistant for Creating an Innovative
  E-commerce Experience}.
\newblock In \emph{CIKM}, 2017.

\bibitem[Liu et~al.(2015)Liu, Gao, He, Deng, Duh, and
  Wang]{Liu2015RepresentationLU}
X.~Liu, J.~Gao, X.~He, L.~Deng, K.~Duh, and Y.-Y. Wang.
\newblock {Representation Learning Using Multi-Task Deep Neural Networks for
  Semantic Classification and Information Retrieval}.
\newblock In \emph{HLT-NAACL}, 2015.

\bibitem[Liu et~al.(2019)Liu, He, Chen, and Gao]{Liu2019MultiTaskDN}
X.~Liu, P.~He, W.~Chen, and J.~Gao.
\newblock {Multi-Task Deep Neural Networks for Natural Language Understanding}.
\newblock \emph{CoRR}, abs/1901.11504, 2019.

\bibitem[Nguyen et~al.(2016)Nguyen, Rosenberg, Song, Gao, Tiwary, Majumder, and
  Deng]{Marco}
T.~Nguyen, M.~Rosenberg, X.~Song, J.~Gao, S.~Tiwary, R.~Majumder, and L.~Deng.
\newblock {MS MARCO: A Human Generated MAchine Reading COmprehension Dataset}.
\newblock \emph{CoRR}, abs/1611.09268, 2016.

\bibitem[Oddy(1977)]{Oddy1977Information}
R.~N. Oddy.
\newblock \emph{{Information Retrieval through Man-Machine Dialogue.}}
\newblock 1977.

\bibitem[Peters et~al.(2018)Peters, Neumann, Iyyer, Gardner, Clark, Lee, and
  Zettlemoyer]{Peters2018DeepCW}
M.~E. Peters, M.~Neumann, M.~Iyyer, M.~Gardner, C.~Clark, K.~Lee, and L.~S.
  Zettlemoyer.
\newblock {Deep contextualized word representations}.
\newblock In \emph{NAACL-HLT}, 2018.

\bibitem[Qu et~al.(2018)Qu, Yang, Croft, Trippas, Zhang, and
  Qiu]{Qu2018AnalyzingAC}
C.~Qu, L.~Yang, W.~B. Croft, J.~R. Trippas, Y.~Zhang, and M.~Qiu.
\newblock {Analyzing and Characterizing User Intent in Information-seeking
  Conversations}.
\newblock In \emph{SIGIR}, 2018.

\bibitem[Qu et~al.(2019{\natexlab{a}})Qu, Yang, Croft, Scholer, and
  Zhang]{answer_interaction}
C.~Qu, L.~Yang, W.~B. Croft, F.~Scholer, and Y.~Zhang.
\newblock {Answer Interaction in Non-factoid Question Answering Systems}.
\newblock In \emph{CHIIR}, 2019{\natexlab{a}}.

\bibitem[Qu et~al.(2019{\natexlab{b}})Qu, Yang, Croft, Zhang, Trippas, and
  Qiu]{UserIntentPred}
C.~Qu, L.~Yang, W.~B. Croft, Y.~Zhang, J.~R. Trippas, and M.~Qiu.
\newblock {User Intent Prediction in Information-seeking Conversations}.
\newblock In \emph{CHIIR}, 2019{\natexlab{b}}.

\bibitem[Qu et~al.(2019{\natexlab{c}})Qu, Yang, Qiu, Croft, Zhang, and
  Iyyer]{hae}
C.~Qu, L.~Yang, M.~Qiu, W.~B. Croft, Y.~Zhang, and M.~Iyyer.
\newblock {BERT with History Answer Embedding for Conversational Question
  Answering}.
\newblock \emph{CoRR}, abs/1905.05412, 2019{\natexlab{c}}.

\bibitem[Rajpurkar et~al.(2016)Rajpurkar, Zhang, Lopyrev, and Liang]{squad}
P.~Rajpurkar, J.~Zhang, K.~Lopyrev, and P.~Liang.
\newblock {SQuAD: 100, 000+ Questions for Machine Comprehension of Text}.
\newblock In \emph{EMNLP}, 2016.

\bibitem[Rajpurkar et~al.(2018)Rajpurkar, Jia, and Liang]{squad2}
P.~Rajpurkar, R.~Jia, and P.~Liang.
\newblock {Know What You Don't Know: Unanswerable Questions for SQuAD}.
\newblock In \emph{ACL}, 2018.

\bibitem[Reddy et~al.(2018)Reddy, Chen, and Manning]{coqa}
S.~Reddy, D.~Chen, and C.~D. Manning.
\newblock {CoQA: A Conversational Question Answering Challenge}.
\newblock \emph{CoRR}, abs/1808.07042, 2018.

\bibitem[Seo et~al.(2016)Seo, Kembhavi, Farhadi, and Hajishirzi]{bidaf}
M.~J. Seo, A.~Kembhavi, A.~Farhadi, and H.~Hajishirzi.
\newblock {Bidirectional Attention Flow for Machine Comprehension}.
\newblock \emph{CoRR}, abs/1611.01603, 2016.

\bibitem[Thomas et~al.(2017)Thomas, McDuff, Czerwinski, and Craswell]{misc}
P.~Thomas, D.~McDuff, M.~Czerwinski, and N.~Craswell.
\newblock {MISC: A data set of information-seeking conversations}.
\newblock In \emph{SIGIR (CAIR'17)}, 2017.

\bibitem[Trippas et~al.(2018)Trippas, Spina, Cavedon, Joho, and
  Sanderson]{Trippas2018InformingTD}
J.~R. Trippas, D.~Spina, L.~Cavedon, H.~Joho, and M.~Sanderson.
\newblock {Informing the Design of Spoken Conversational Search: Perspective
  Paper}.
\newblock In \emph{CHIIR}, 2018.

\bibitem[Vaswani et~al.(2017)Vaswani, Shazeer, Parmar, Uszkoreit, Jones, Gomez,
  Kaiser, and Polosukhin]{transformer}
A.~Vaswani, N.~Shazeer, N.~Parmar, J.~Uszkoreit, L.~Jones, A.~N. Gomez,
  L.~Kaiser, and I.~Polosukhin.
\newblock {Attention Is All You Need}.
\newblock In \emph{NIPS}, 2017.

\bibitem[Wang et~al.(2017)Wang, Yang, Wei, Chang, and Zhou]{Wang2017GatedSN}
W.~Wang, N.~Yang, F.~Wei, B.~Chang, and M.~Zhou.
\newblock {Gated Self-Matching Networks for Reading Comprehension and Question
  Answering}.
\newblock In \emph{ACL}, 2017.

\bibitem[Xu et~al.(2018)Xu, Liu, Shen, Liu, and Gao]{Xu2018MultiTaskLF}
Y.~Xu, X.~Liu, Y.~Shen, J.~Liu, and J.~Gao.
\newblock {Multi-Task Learning for Machine Reading Comprehension}.
\newblock \emph{CoRR}, abs/1809.06963, 2018.

\bibitem[Yang et~al.(2017)Yang, Zamani, Zhang, Guo, and
  Croft]{Yang2017NeuralMM}
L.~Yang, H.~Zamani, Y.~Zhang, J.~Guo, and W.~B. Croft.
\newblock {Neural Matching Models for Question Retrieval and Next Question
  Prediction in Conversation}.
\newblock \emph{CoRR}, 2017.

\bibitem[Yang et~al.(2018)Yang, Qiu, Qu, Guo, Zhang, Croft, Huang, and
  Chen]{Yang2018ResponseRW}
L.~Yang, M.~Qiu, C.~Qu, J.~Guo, Y.~Zhang, W.~B. Croft, J.~Huang, and H.~Chen.
\newblock {Response Ranking with Deep Matching Networks and External Knowledge
  in Information-seeking Conversation Systems}.
\newblock In \emph{SIGIR}, 2018.

\bibitem[Yatskar(2018)]{Yatskar2018AQC}
M.~Yatskar.
\newblock {A Qualitative Comparison of CoQA, SQuAD 2.0 and QuAC}.
\newblock \emph{CoRR}, abs/1809.10735, 2018.

\bibitem[Zhang and Yang(2018)]{Zhang2018ASO}
Y.~Zhang and Q.~Yang.
\newblock {A Survey on MultiTask Learning}.
\newblock 2018.

\bibitem[Zhang et~al.(2018)Zhang, Chen, Ai, Yang, and
  Croft]{Zhang2018TowardsCS}
Y.~Zhang, X.~Chen, Q.~Ai, L.~Yang, and W.~B. Croft.
\newblock {Towards Conversational Search and Recommendation: System Ask, User
  Respond}.
\newblock In \emph{CIKM}, 2018.

\bibitem[Zhu et~al.(2018)Zhu, Zeng, and Huang]{sdnet}
C.~Zhu, M.~Zeng, and X.~Huang.
\newblock {SDNet: Contextualized Attention-based Deep Network for
  Conversational Question Answering}.
\newblock \emph{CoRR}, 2018.

\end{thebibliography}
